\documentclass[print, aps, amssymb, prl, superscriptaddress, twocolumn]{revtex4}  
\usepackage[draft]{hyperref} 
\usepackage{amsmath}
\usepackage{amssymb}
\usepackage{amsthm}
\usepackage{wasysym}
\usepackage{graphicx}

\begin{document}

\title{All-optical Switching of a Microcavity \\ Repeated at Terahertz Rates}

\author{Emre Y\"uce,$^{1*}$ Georgios Ctistis,$^{1}$ Julien Claudon,$^{2}$ Emmanuel Dupuy,$^{2}$ Robin D. Buijs,$^{1}$ Bob de Ronde,$^{1}$ Allard P. Mosk,$^{1}$ Jean-Michel G\'erard,$^{2}$ Willem L. Vos$^{}$}
\affiliation{$^{}$Complex Photonic Systems (COPS), MESA+ Institute for
Nanotechnology, University of Twente, P.O. Box 217, 7500 AE Enschede, The Netherlands \\
$^{2}$CEA-CNRS-UJF ``Nanophysics and Semiconductors" Joint
Laboratory, CEA/INAC/SP2M, 17 rue des Martyrs, 38054 Grenoble
C\'edex 9, France\\
$^*$Corresponding author: e.yuce@utwente.nl: www.photonicbandgaps.com
}

\begin{abstract}We have performed ultrafast pump-probe experiments on a GaAs-AlAs microcavity with a resonance near 1300 nm in the ``original" telecom band. We exploit the virtually instantaneous electronic Kerr effect to repeatedly and reproducibly switch a GaAs-AlAs planar microcavity. We achieve repetition times as fast as 300 fs, thereby breaking the THz modulation barrier. The rate of the switching in our experiments is only determined by optics and not by material related relaxation. Our results offer novel opportunities for fundamental studies of cavity-QED and optical information processing in sub-picosecond time scale.\end{abstract}

\maketitle

\noindent Semiconductor microcavities have proven to be essential to strongly confine light, thereby enhancing the interaction between light and matter~\cite{gerard.1998.prl} to the point of manipulating quantum states of matter~\cite{reithmaier.2004.nature, khitrova.2006.nat.phys, englund.2012.prl}. To achieve dynamic control of these processes, switching speed must approach their characteristic time scales~\cite{gerard.1998.prl, khitrova.2006.nat.phys} in the picosecond range, corresponding to THz modulation rates. For real-time cavity-QED, the switching time should be shorter than relevant time-scales such as the lifetime of an emitter in the weak coupling regime $\tau_{qd}$, or the period of the Rabi-oscillation $\tau_{Rabi}$ of a two-level system in the strong coupling regime. For a single quantum dot in a semiconductor microcavity, such times are of the order of $\tau_{qd}=20-200 \ \rm{ps}$ and $\tau_{Rabi}=1-10 \ \rm{ps}$~\cite{gerard.1998.prl, khitrova.2006.nat.phys}. Thus switching on the ps time scale or faster is essential. While all-optical switching is essential as it allows sub-picosecond reversible control of the refractive index~\cite{boyd.1992, hulin.1986.apl, georgios.2011.apl}, the popular optical free carrier excitation mechanisms~\cite{leonard.2002.prb, almeida.2004.ol, harding.2007.apl, nozaki.2010.nat.ph, song.2010.ieeetqe} are too slow to repeatedly switch at THz rates~\cite{wada.2004.njphys}. 

We exploit the electronic Kerr effect that enables switching of the cavity faster than all relevant time scales. We employ the virtually instantaneous electronic Kerr effect to repeatedly and reproducibly switch a GaAs-AlAs planar microcavity operating in the telecom O-band. 


\begin{figure}[htb]
  \begin{center}
  \includegraphics[width=8.3 cm]{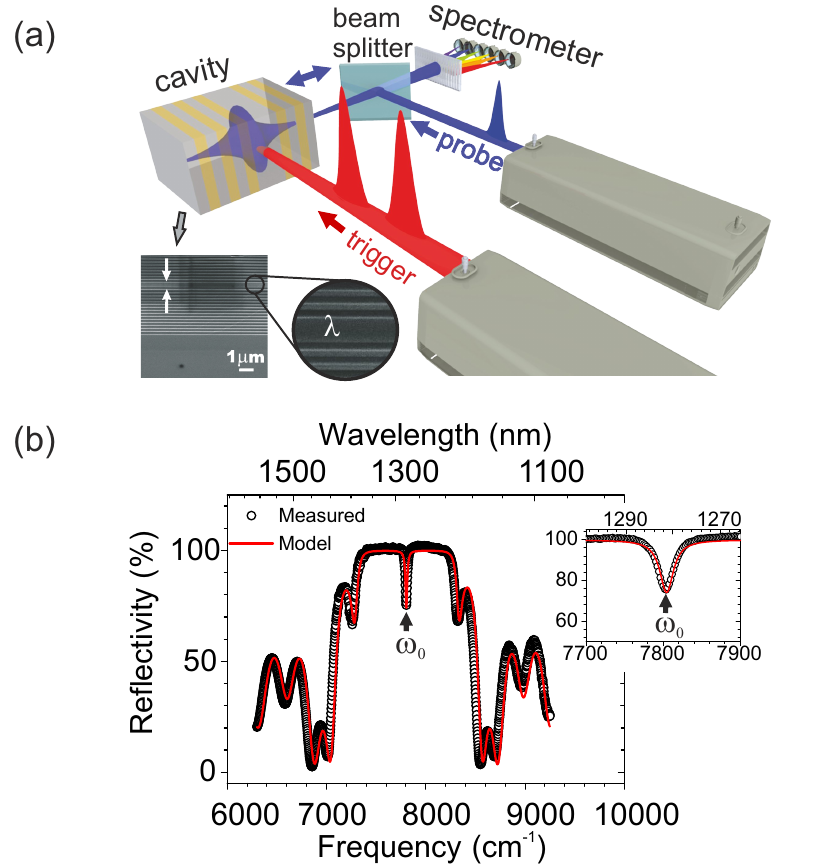}
  \caption{(colour) \emph{(a) Schematic of the set-up. The probe beam path is shown in blue, the trigger beam path with multiple pulses in red. Signal reflected from the cavity is spectrally resolved and detected with a spectrometer. Inset: Scanning electron micrograph of the microcavity with the $\lambda$-thick defect layer (arrows) sandwiched between two Bragg stacks. (b) Measured (black circles) and calculated (transverse matrix model red curve) reflectivity spectra of the microcavity. The stopband of the Bragg stacks extends from $7072 \ \rm{cm^{-1}}$ to $8498 \ \rm{cm^{-1}}$.  A narrow trough at $\omega_0=7804 \ cm^{-1}$ ($\lambda_{0} = 1281.4 \ nm$) indicates the cavity resonance, also shown in the inset with a high resolution.}}
\label{yuce_f1}
\end{center}
\end{figure}

Physically, the electronic Kerr effect is one of the fastest optical phenomena on account of the low mass of electrons. In many practical situations, however, slower nonlinear effects such as two-photon excited free carriers dominate over the electronic Kerr effect~\cite{boyd.1992, yuce.2012.josab}. In order to avoid two-photon absorption we designed our experiment to operate with low photon-energy trigger pulses. We employ two independently tunable lasers as sources of trigger and probe with pulse durations of $\tau_{P} = 140 \pm 10 \ \rm{fs}$ as shown in Fig.~\ref{yuce_f1}(a). The frequency ($\omega_{tr}$) of the trigger light is centred at $\omega_{tr}= 4165 \ \rm{cm^{-1}}$ $(\lambda_{tr}=2400 \rm{nm})$ to suppress non-degenerate two photon absorption ($\omega_{pr}+\omega_{tr}\leq \omega_{gap}$)~\cite{yuce.2012.josab}, and to avoid degenerate two photon absorption ($2\omega_{tr}<\omega_{gap}$). The trigger pulse energy is set to $I_{tr}=22 \pm 2 \rm{\ pJ / \mu m^2}$. The trigger beam has a larger Gaussian focus ($\diameter _{tr}= 70 \ \rm{\mu m}$) than the probe ($\diameter _{pr}=30 \ \rm{\mu m}$). Michelson interferometers are used in both trigger and probe beam paths to split each pulse into two pulses. The time delay between the successive trigger ($\delta t_{tr}$) and probe ($\delta t_{pr}$) pulses is set to be in the range of 1 ps, which corresponds to THz repetition rates. We perform our experiments at room temperature.   

The sample consists of a GaAs $\lambda$-layer ($d=376 \ \rm{nm}$) sandwiched between two Bragg stacks made of 7 and 19 pairs of $\lambda/4$-thick layers of GaAs ($d_{GaAs}=94 \ \rm{nm}$) and AlAs ($d_{AlAs}= 110 \ \rm{nm}$), respectively. The relatively high reflectivity of the resonance minimum ($R_{trough}=80\ \%$) is caused by the asymmetric cavity design (Fig.~\ref{yuce_f1}(b)). From the linewidth ($\Delta \omega=20  \pm 3 \ \rm{cm^{-1}}$) of the cavity resonance we derive a quality factor $Q=390 \pm 60$ and a cavity storage time $\tau_{cav}=300 \pm 50 \ \rm{fs}$. We take the resonance frequency of the cavity to be the minimum of the cavity trough, see Fig.~\ref{yuce_f1}(b). We deliberately reduced the storage time of the probe photons in the cavity by decreasing the reflectivity of the top mirror. This allows for fast switching rates, and simultaneously multi-photon excitation of free carriers is reduced compared to high Q cavities~\cite{yuce.2012.josab}. 


\begin{figure}[htb]
  \begin{center}
  \includegraphics[width=8.3 cm]{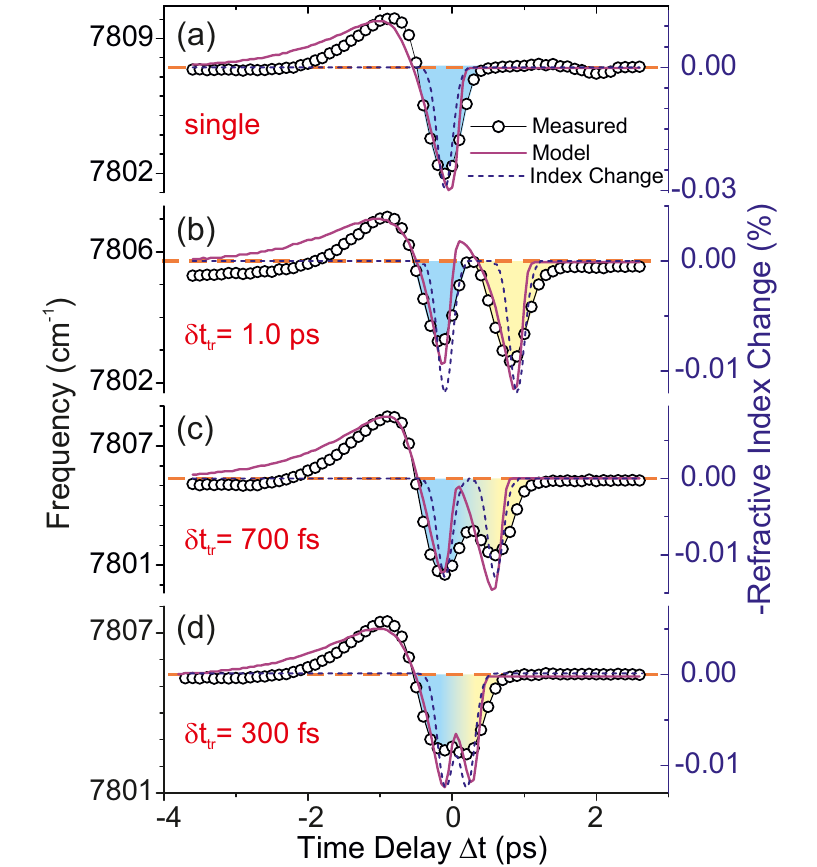}
  \caption{(colour) \emph{Cavity resonance frequency and refractive index change versus time delay $\Delta t$. (a) One trigger pulse is applied. (b) Two trigger pulses separated by $\delta t_{tr}$=1.0 ps, (c) $\delta t_{tr}$=700 fs, and (d) $\delta t_{tr}$=300 fs. The horizontal dashed lines denote the unswitched resonance frequency. The solid curves represent the calculated resonance frequency and the dotted lines the refractive index change in the dynamic model. Blue-filled regions indicate the overlap of the first trigger pulse with the probe pulse and the yellow regions indicate the second trigger pulse.}}
  \label{yuce_f2}
  \end{center}
\end{figure}

Figure~\ref{yuce_f2} shows the resonance frequency versus trigger-probe time delay $\Delta t$ for different separations of two trigger pulses ($\delta t_{tr}$). We start in Fig.~\ref{yuce_f2}(a) with widely separated trigger pulses, so that single switch events are well separated. The resonance quickly shifts by $5.7 \ \rm{cm^{-1}}$ to a lower frequency at trigger-probe overlap ($\Delta t=0$) and returns to the starting frequency immediately after the trigger pulse is gone. Single sub-picosecond switching events have also been observed for excited quantum wells in a planar microcavity~\cite{hulin.1986.apl}. In contrast, the electronic Kerr effect offers more flexibility as it is not restricted to a specific resonance and geometry. We conclude from the shift to a lower frequency that the refractive index of GaAs has increased, in agreement with the positive non-degenerate Kerr coefficient of GaAs~\cite{yuce.2012.josab}. To confirm our hypothesis, we performed calculations with a dynamic model, taking solely into account the positive refractive index change of GaAs due to the electronic Kerr effect. The propagation of the field through the cavity is calculated by evaluating the incident fields at each of the interfaces in the multilayer in the time domain~\cite{harding.2012.josab}. Figure~\ref{yuce_f2} shows that the minimum of the resonance trough appears at a higher frequency when the probe pulse arrives before the trigger pulse ($\Delta t<-500$ fs) while the index of the cavity only increases during the overlap of trigger-probe pulses. The apparent blue-shift of the minimum of the cavity trough is due to the interference between light that is directly reflected from the front Bragg mirror and light that escapes from the cavity. Nonetheless, the refractive index has not yet changed at $\Delta t<-500$ fs. We observe an excellent quantitative agreement between the measured and calculated shifts for the magnitude of the shift and the temporal evolution.

Figure~\ref{yuce_f2} shows the ultrafast control of the microcavity resonance under the influence of repeated trigger pulses that we generated through the trigger pulse divider. The delay of the second trigger pulse with respect to the first pulse is set to (b) $\delta t_{tr}=$ 1.0 ps, (c) $\delta t_{tr}=$ 700 fs, and (d) $\delta t_{tr}=$ 300 fs. Figures~\ref{yuce_f2}(b-d) show that the measured resonance frequency shifts to lower frequencies at $\Delta t=0$ due to the first trigger pulse. Right after ($\Delta t=\delta t_{tr}$) we reversibly change the resonance frequency again with the second trigger pulse. At a trigger repetition time of $\delta t_{tr}=300 \ \rm{fs}$, the consecutive switch events are barely resolved since the fundamental speed limit is set by the cavity storage time $\tau_{cav}=300 \ \rm{fs}$ in our study. The fact that after the trigger pulse the resonance frequency returns to its initial value within 300 fs confirms the absence of free carrier effects. The relatively slow decay of excited free carriers over tens of picoseconds or more would indeed impede the repeated switching of the cavity on a sub-picosecond time scale~\cite{leonard.2002.prb, harding.2007.apl}. The very good agreement between the calculation and experimental results confirms the ultimate fast electronic Kerr switching of our photonic microcavity at THz repetition rate. The repetition rate achieved here is already an order of magnitude larger than the fastest reported cavity modulators in the telecom range~\cite{almeida.2004.ol, gil.2009.ieee, nozaki.2010.nat.ph}.
 
\begin{figure}[htb]
  \begin{center}
  \includegraphics[width=8.3 cm]{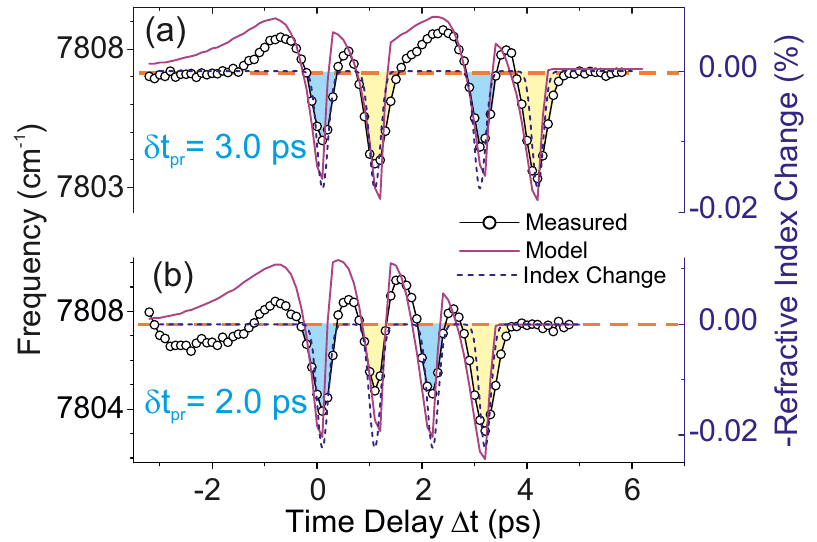}
  \caption{(colour) \emph{Cavity resonance frequency and refractive index change versus time delay $\Delta t$. The cavity resonance frequency is switched with two trigger pulses separated by $\delta t_{tr}$=1.0 ps and two probe pulses separated by (a) $\delta t_{pr}$=3.0 ps (b) $\delta t_{pr}$=2.0 ps. Horizontal dashed lines are the unswitched resonance. Solid curves are the calculated resonance frequency and the dotted lines show the refractive index change in the dynamic model. Blue-filled regions indicate the overlap of the first trigger pulse with probe pulses and yellow regions indicate the overlap of the second trigger pulse with the probe pulses.}}
  \label{yuce_f3}
  \end{center}
\end{figure}

Figure~\ref{yuce_f3} shows the resonance frequency versus trigger-probe time delay $\Delta t$ for different delays $\delta t_{pr}$ of two probe pulses at a fixed time delay $\delta t_{tr}=1 \ \rm{ps}$ between successive trigger pulses. In this experiment, we have used two closely timed probe pulses to monitor twice each of the two switching events separated by $\delta t_{tr}=1$ ps. As expected, four ultrafast switching events are observed as a function of the trigger-probe delay. Each pair of trigger pulses interacts with each probe pulse resulting in instantaneous refractive index change. Figure~\ref{yuce_f3}(a) and (b) show that all four switch events are reversible, similar, and not affected by the prior switching events as little as 1 ps earlier. Therefore, we demonstrate repeated triggering and probing of the system at THz repetition rates.


Using our method, the switching speed and the repetition rate can be controlled for a specific application by choosing the storage time of the cavity and the timing between successive trigger pulses. The recent developments in the compound ultrafast laser systems~\cite{wada.2004.njphys, oldenbeuving.2010.oe} promise high repetition rates that are beneficial for repeated switching in picosecond time scale. From the measured nondegenerate two- and three-photon absorption coefficients~\cite{yuce.2012.josab}, we estimate the absorbed fraction of the trigger pulse to be as small as $10^{-6}$ ($a\leq0.08 \ \rm{fJ/\mu m^2}$). Unlike the free carrier excitation schemes we here avoid absorption so that the trigger photons could be recycled to switch the cavity again. Switching with the Kerr effect at even lower pulse energies~\cite{matsuda.2009.nat.photon} is achieved if the trigger pulses are also resonantly enhanced by the cavity. Our results offer novel opportunities for fundamental studies of cavity-QED, frequency conversion and optical information processing in sub-picosecond time scale.




\end{document}